\documentclass[aps,pra,twocolumn,floatfix,nofootinbib,longbibliograpphy]{revtex4-2}
\usepackage{dcolumn}
\usepackage{bm}
\usepackage{graphicx,subfigure,xcolor}
\usepackage{braket}
\usepackage[normalem]{ulem}
\usepackage{balance}
\usepackage{comment}
\usepackage{xcolor}
\usepackage{amsmath}
\usepackage{amssymb}
\usepackage{eucal}
\usepackage{mathrsfs}
\usepackage{amsthm}
\usepackage{epstopdf}
\usepackage{textcomp}
\usepackage{braket}
\usepackage[utf8]{inputenc}
\usepackage[T1]{fontenc}
\usepackage{xcolor}
\usepackage{soul}

\newcommand{\wo}{\omega_0}
\newcommand{\wj}{{\omega_j}}
\newcommand{\wk}{{\omega_k}}
\newcommand{\wl}{{\omega_\ell}}
\newcommand{\wm}{{\omega_m}}
\newcommand{\wpl}{{\omega_{pl}}}
\newcommand{\wq}{{\omega_q}}
\newcommand{\ws}{{\omega_s}}

\newcommand{\Lo}{L_0}

\begin{document}

\title{Spatial correlations of field observables in two half-spaces separated by a movable perfect mirror}

\author{Federico Montalbano$^{1}$}
\author{Federico Armata$^{1}$}
\author{Lucia Rizzuto$^{1,2}$}
\author{Roberto Passante$^{1,2}$}
\email{roberto.passante@unipa.it}

\affiliation{$^1$ Dipartimento di Fisica e Chimica - Emilio Segr\`{e},
Universit\`{a} degli Studi di Palermo, Via Archirafi 36, I-90123 Palermo,
Italy}
\affiliation{$^2$ INFN, Laboratori Nazionali del Sud, I-95123 Catania, Italy}

\begin{abstract}
We consider a system of two cavities separated by a reflecting boundary of finite mass that is free to move, and bounded to its equilibrium position by a harmonic potential. This yields an effective mirror-field interaction, as well as an effective interaction between the field modes mediated by the movable boundary. Two massless scalar fields are defined in each cavity. We consider the second-order interacting ground state of the system, that contains virtual excitations of both mirror's degrees of freedom and of the scalar fields. We investigate the correlation functions between field observables in the two cavities, and find that the squared scalar fields in the two cavities, in the interacting ground state, are anticorrelated. We discuss the dependence of the correlation on the distance of the two points considered from the mirror's average position and on its mass and oscillation angular frequency. These results show a sort of communication between the two half-spaces separated by the movable mirror, mediated by its position fluctuations. Observability of this new phenomenon exploiting two- or many-body dispersion interactions between polarizable bodies is discussed. The dependence on a cutoff frequency introduced to regularize the frequency integrations, as well as the case of a real conductor, are also discussed.
\end{abstract}

\maketitle

\section{\label{sec:1}Introduction}

In quantum field theory, the presence of a dielectric or reflecting boundary affects the structure of the field modes and can change the expectation values of several local field observables, for example the field energy density, spatial field correlations or dispersion interactions \cite{Milonni94,Buhmann15,Milton11,Bartolo-Passante12}. This has relevant observable consequences, for example changes in the spontaneous emission rate of an atom and atom-surface Casimir-Polder interactions \cite{Milonni94,Milonni19,Messina-Passante08}. In the case of an infinite flat, perfectly reflecting and static mirror, it is well known that the boundary separates the space into two independent half-spaces. In such cases, the mirror is usually treated as an assigned boundary condition on the field operators. In this paper, we will show that the situation significantly changes if the boundary has a finite mass and it is allowed to move, because correlations between field quantities evaluated in the two half-spaces exist. Inclusion of the motion of a boundary is at the basis of quantum optomechanics \cite{Meystre13,Aspelmeyer-Kippenberg14} and dynamical Casimir effect \cite{Moore70,Dodonov10,Dodonov20}. In the context of the dynamical Casimir effect, the mirror's motion is prescribed from the outside, and it is usually included in the formalism in the form of a time-dependent boundary condition on the relevant field operators; the mechanical degrees of freedom are not dynamical variables of the system. On the contrary, another type of systems investigated in the literature, for example in quantum optomechanics, treats the mirror as a quantum system with its mechanical degrees of freedom \cite{Law95}, thus allowing typical quantum effects such as quantum fluctuations of its position. As we shall discuss below, in this paper we adopt this second point of view by including the wall's mechanical degrees of freedom in the Hamiltonian description of our system. A model Hamiltonian describing a quantum field in the presence of a movable boundary of finite mass, bound to its equilibrium position by a harmonic potential, has been introduced in \cite{Law94,Law95}. This effective Hamiltonian has been used to evaluate the field energy density change near the fluctuating boundary, as well as corrections to the Casimir force \cite{Butera-Passante13,Armata-Passante15} and radiation pressure effects on a two-mirror system \cite{Butera22}. These quantum effects become more and more relevant the smaller the mirror mass is \cite{Butera-Passante13}, and in modern quantum optomechanical experiments it can be reduced even to values of the order of $10^{-21}$ Kg \cite{Aspelmeyer-Kippenberg14,Barzanjeh-Xuereb22}.
Also, the changes of the field energy density, with respect to the fixed-wall case, are larger in the very proximity of the movable wall since the virtual quanta emitted by the movable wall are localized near its position due to the energy-time uncertainty relation \cite{Butera-Passante13}.
Internal degrees of freedom of a boundary have been considered also in the case of a moving dielectric membrane \cite{Cheung-Law11}, within microscopic models of the boundary \cite{Galley-Behunin13,Sinha-Lin15} and in connection with dissipative dynamics of a movable particle coupled to a field \cite{Sinha-Lopez21}.

In this paper, we consider a system of two cavities separated by a movable perfectly reflecting wall of a finite mass and bound to its equilibrium position by a harmonic potential. Two massless one-dimensional scalar fields are considered in the two cavities, each of them interacting with the movable mirror separating the two cavities.
Even if our model is one-dimensional, we expect that the qualitative features of the results obtained should be representative of what happens in a more realistic threedimensional case, similarly to the case of the field energy density studied in \cite{Armata-Passante15} for the 3D scalar field. We use the names cavity and wall also for our one-dimensional case, following a common use in the relevant literature.
Our interest is investigating how the fluctuating motion of the movable mirror affects local field properties in the two cavities in the ground state of the system. In \cite{Armata-Kim17} we considered the dynamical self-dressing of the movable wall starting from a nonequilibrium state, in particular the time-dependent energy shift. We found that, up to second order in the wall-field interaction, the self-dressing process is the same of the single-cavity case and thus the two cavities do not influence each other at second order; we argued that the two half-spaces influence each other starting from fourth order in the coupling; a change of the Casimir force between the two fixed walls, mediated by the movable wall, is therefore expected at the fourth order. Here we address a different problem; specifically, we investigate the correlation functions between two field observables in the two cavities. We find that they can start also from the second order, showing a sort of influence (correlation) between the two half-spaces mediated by the perfectly reflecting movable wall. We also discuss how this new effect could be observed exploiting Casimir-Polder interactions between polarizable bodies.

We first generalize the mirror-field interaction Hamiltonian mentioned above and first introduced in \cite{Law95} to the present two-cavity case, and then evaluate by second-order perturbation theory the interacting ground state of the system, that contains virtual excitations of the fields in the cavities and of the mirror (phonons). Successively, we compute expectation values on the ground state of relevant correlation functions of local field observables in the two cavities.
In the frequency integrations, we introduce an appropriate exponential regularization function in order to cure ultraviolet divergences.
We find that the correlation between the scalar fields in the two cavities vanishes, and indeed this happens at any order in the wall-field coupling, within the limits of our Hamiltonian model (essentially in the hypothesis of small displacements of the movable wall). We thus evaluate other correlation functions between field observables in the two cavities.

We find that a nonvanishing spatial anticorrelation between the squared fields in the two cavities exists and explicitly evaluate its distance dependence. This relevant feature is not present in the case of a fixed wall, and sharply distinguishes our trembling wall case from the usual fixed-wall case, even for ideal mirrors.
The squared field operator is an important observable, since the field energy density is for example obtained from the square of the field and of its derivatives.
We discuss the dependence of such correlation from the relevant physical parameters, in particular mass and binding frequency of the movable wall, distance of the two points from the movable wall, as well observability of this phenomenon through retarded Casimir-Polder interactions between two polarizable bodies placed at the two sides of the movable wall.
We also show that for the relevant distance scales considered (i.e. larger than the wavelength associated to the cutoff frequency), our results are quite independent from the cutoff frequency, and that the contribution of high-frequency modes is negligible. We also consider, in the case of a movable real wall made of a metal with a finite plasma frequency, the effect on the correlation functions considered due to the high-frequency modes delocalized among the two half-spaces.
Possible relevance of our findings on correlations of observables evaluated at the two sides of a fluctuating event horizon (as supposed in quantum gravity theories) is also mentioned.

This paper is organized as follows. In Sec. \ref{sec:2} we introduce our two-cavity system and its Hamiltonian, describing an effective wall-field interaction, and obtain by perturbation theory the interacting ground state of the system. In Sec. \ref{sec:3} we evaluate the expectation value on this state of relevant field quantities, specifically the spatial correlation function between points in the two cavities of the squared field; we find that an anticorrelation exists and we discuss its main physical features. Section \ref{sec:4} is devoted to our conclusive remarks.

\section{\label{sec:2}The two-cavity system}

\begin{figure}[!htbp]
\centering\includegraphics[width=8.6cm]{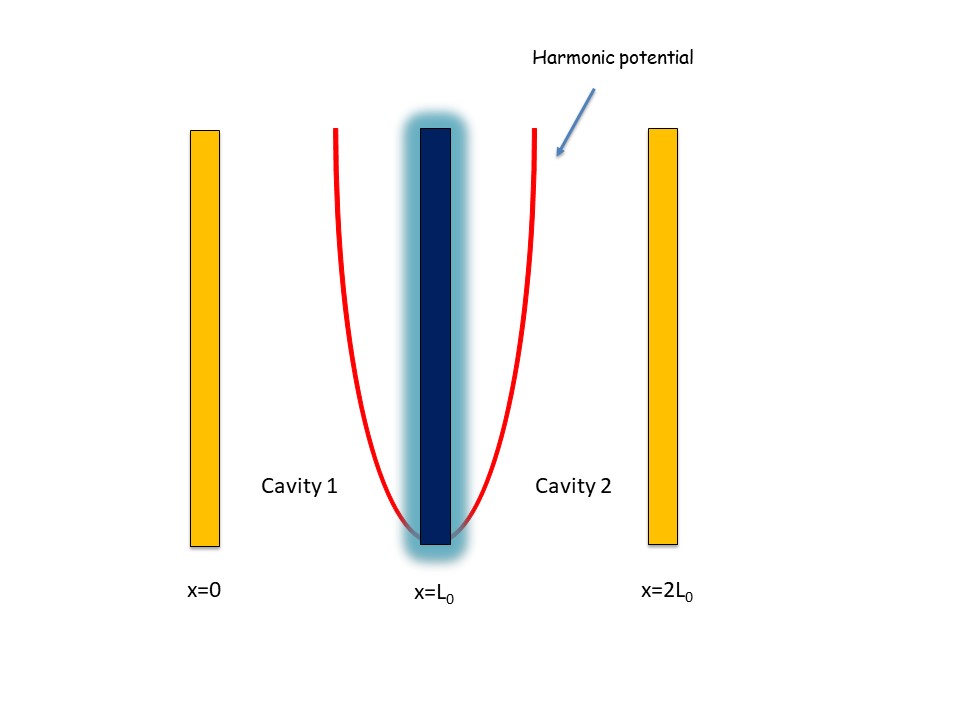}
\caption{The two-cavity system: two fixed ideal mirrors at $x=0$ and $x=2\Lo$, separated by an ideal movable mirror with its equilibrium position at $x=\Lo$. In this paper, we consider a one-dimensional case for the two cavities.}
\label{Fig1}
\end{figure}

Our physical system is constituted by two ideal cavities, separated by a movable perfectly conducting mirror of mass $m$, and bounded to its equilibrium position $x=\Lo$ by a harmonic potential of frequency $\wo$. The two fixed mirrors have positions $x=0$ and $x=2\Lo$. Two massless scalar quantum fields are defined in the two half-spaces separated by the movable mirror. For simplicity, we here consider a one-dimensional model. The situation is shown in Fig. \ref{Fig1}. The mechanical degrees of freedom of the movable mirror are treated quantum mechanically, and the mirror is described as a quantum harmonic oscillator of frequency $\wo$; we label with $b$ and $b^\dagger$ its annihilation and creation operators, respectively. Also, we respectively label with $a_p$ and $a_p^\dagger$ the bosonic annihilation and creation operators of the massless scalar field on the left cavity ($0<x<\Lo$), and with $c_r$ and $c_r^\dagger$ those for the cavity on the right ($\Lo <x<2\Lo$); all these operators refer to the field modes relative to the equilibrium position $\Lo$ of the movable wall. For a single cavity, and in the hypothesis of small displacement of the movable wall from its equilibrium position, the system can be described on the basis of an effective interaction. This Hamiltonian was first introduced by Law for a single cavity with a movable mirror, in terms of field modes relative to the equilibrium position of the movable boundary \cite{Law95,Law94}. The Law Hamiltonian, describing the effective mirror-field interaction, is linear in the mirror coordinates and bilinear in the field coordinates,
and yields an effective wall-field interaction, as well as an effective interaction between the field modes, mediated by the movable mirror.
This Hamiltonian follows from a field quantization with a movable wall, bounded to an equilibrium position, assuming that it undergoes small displacements, and it is completely expressed in terms of bosonic operators relative to its equilibrium position. It was originally developed for the one-dimensional case \cite{Law95}, but it can be easily generalized to the three dimensional case \cite{Armata-Passante15}. In order to avoid cumbersome calculations involving discrete summations over the modes in the three dimensions, not essential for our aims, we limit the present investigation to the one-dimensional case only, that already contains all the relevant elements we are interested in.

Our first step is to generalize the Law Hamiltonian to our two-cavity scenario. This is quite immediate: we quantize the two scalar fields in the two half-spaces with respect to the equilibrium position $x=\Lo$ of the mobile wall, and take into account that if a movable wall's displacement in one direction makes shorter one cavity, it makes longer the other cavity. For some general preliminary considerations of this two-cavity system, see also \cite{Armata-Kim17}. Our Hamiltonian including the effective interaction of the movable wall with the two scalar fields is then of the form $H=H_0+H_I$, with the unperturbed term
\begin{equation}
\label{H0}
H_0 = \hbar \wo b^\dagger b + \hbar \sum _k \wk a_k^\dagger a_k + \hbar \sum _k \wk c_k^\dagger c_k ,
\end{equation}
and the interaction term $H_I=H_I^1 + H_I^2$, where $H_I^1$ and $H_I^2$ are, respectively, the effective interaction Hamiltonians between the movable wall and the scalar field in the left (1) and right (2) half-spaces. These interaction terms are given by
\begin{eqnarray}
\label{effint}
H_I^1 &=& -\left( b+b^\dagger \right) \sum_{kj} C_{kj}^1 \text{N} \left[ \left( a_j + a_j^\dagger \right) \left( a_k + a_k^\dagger \right) \right] ,
\nonumber \\
H_I^2 &=& -\left( b+b^\dagger \right) \sum_{kj} C_{kj}^2 \text{N} \left[ \left( c_j + c_j^\dagger \right) \left( c_k + c_k^\dagger \right) \right] ,
\end{eqnarray}
where N is the normal ordering operator, and the coupling constants are given by
\begin{eqnarray}
\label{couplconst}
C_{kj}^1 &=& (-1)^{j+k} \left( \frac {\hbar}2 \right)^{3/2} \frac 1{\Lo \sqrt{m}}\sqrt{\frac {\wj \wk}{\wo}} ,
\\
C_{kj}^2 &=& -C_{kj}^1 ,
\end{eqnarray}
with $\wk = ck$. The coupling constants $C_{kj}^1$ and $C_{kj}^2$ have an opposite sign since, when the wall moves from its equilibrium position, the sign of its displacement is opposite for the two cavities. This Hamiltonian holds for small displacements of the movable wall around its equilibrium position $\Lo$, and it is the immediate generalization to our two-cavity case of the Hamiltonian introduced in \cite{Law95}
and discussed before. It is quadratic in the annihilation and creation operators relative to the two half-spaces, and its structure immediately yields an effective interaction between the field modes (that are independent in the usual case of fixed cavity walls).

A generic eigenstate of the unperturbed Hamiltonian $H_0$ will be labeled as
\begin{equation}
\label{unpeigen}
\lvert n_{wall}; \{ n^1\} ; \{ n^2 \} \rangle ,
\end{equation}
where the first item $n_{wall}$ denotes the number of excitations in the mirror (phonons), while $\{n^1\}$ and $\{n^2\}$ denote, respectively, a complete set of occupation numbers of the field modes of the first and of the second cavity. The unperturbed ground state is thus $\lvert g \rangle =\lvert 0; \{ 0\} ; \{ 0\} \rangle$, with unperturbed energy $E_g=0$.

We can now obtain the first- and second-order corrections to the ground state, due to the mirror-field interaction, by stationary perturbation theory. The first-order correction is
\begin{widetext}
\begin{equation}
\label{corr-1}
\lvert g^{(1)} \rangle = \sum_{jk} \frac {C_{kj}^1}{\hbar ( \wo +\wj +\wk )} \lvert 1; \{ 1_j 1_k\}; \{ 0\} \rangle
+ \sum_{jk} \frac {C_{kj}^2}{\hbar ( \wo +\wj +\wk )} \lvert 1; \{ 0\}; \{ 1_j 1_k\} \rangle ,
\end{equation}
\end{widetext}
which contains states with one mirror excitation and a pair of quanta in one of the two cavities. These are virtual mirror and field excitations, of course.

The second-order correction is given by the sum of several terms, that, for convenience of notation, we arrange in the following form
\begin{widetext}
\begin{eqnarray}
\label{corr-2}
\lvert g^{(2)} \rangle &=& \lvert g^{(2)}_1 \rangle + \lvert g^{(2)}_2 \rangle + \lvert g^{(2)}_{11} \rangle + \lvert g^{(2)}_{2w} \rangle ,
\\
\lvert g^{(2)}_1 \rangle &=&  \sum_{jk\ell} \frac 4{\hbar^2(\wo +\wj +\wk)(\wl +\wj)}
\Big[ C^1_{jk} C^1_{\ell k} \lvert 0; \{ 1_\ell 1_j \}; \{ 0 \} \rangle + C^2_{jk} C^2_{\ell k} \lvert 0; \{ 0 \}; \{ 1_\ell 1_j \} \rangle \Big] ,
\\
\lvert g^{(2)}_2 \rangle &=&  \sum_{jk\ell m} \frac 1{\hbar^2(\wo +\wj +\wk )(\wj +\wk +\wl +\wm )}
\Big[ C^1_{jk} C^1_{\ell m} \lvert 0; \{ 1_j 1_k 1_\ell 1_m \}; \{ 0 \} \rangle
\nonumber \\
&\ & \ + C^2_{jk} C^2_{\ell m} \lvert 0; \{ 0 \}; \{ 1_j 1_k 1_\ell 1_m \} \rangle \Big] ,
\\
\lvert g^{(2)}_{11} \rangle &=&  \sum_{jk\ell m} \left[ \frac { C^1_{jk} C^2_{\ell m}}{\hbar^2(\wo +\wj +\wk)(\wj +\wk +\wl +\wm)} +
\frac { C^1_{jk} C^2_{\ell m}}{\hbar^2(\wo +\wl +\wm )(\wj +\wk +\wl +\wm )} \right]
\nonumber \\
&\ & \ \times \lvert 0; \{ 1_j 1_k \}; \{ 1_\ell 1_m \} \rangle ,
\end{eqnarray}
\end{widetext}
while $\lvert g^{(2)}_{2w} \rangle$ are terms with two mirror's excitations, that do not contribute to the expectation values we will evaluate in the next section, and thus we do not explicitly write them down.

The normalized corrected ground state assumes thus the form
\begin{equation}
\label{corrected state}
\lvert \tilde{g} \rangle = \left( 1 -\frac 12 \Lambda^2 \right)\lvert 0; \{ 0\} ; \{ 0\} \rangle + \lvert g^{(1)} \rangle +  \lvert g^{(2)} \rangle ,
\end{equation}
where $\Lambda$ is a normalization factor obtained, in second-order perturbation theory, from
\begin{equation}
\label{normalization}
\Lambda^2 = \sum_{jk} \frac{(C^1_{jk})^2 + (C^2_{jk})^2}{\hbar^2(\wo +\wj + \wk )^2} .
\end{equation}

In the next section, we will evaluate the expectation value on the corrected ground state (\ref{corrected state}) of correlation functions between field observables evaluated at different points of the two cavities.

\section{\label{sec:3}Correlation functions of field observables in the two cavities}

In this section we evaluate, on the corrected ground state, the correlation functions of field observables in the two half-spaces, and discuss their observability through dispersion interactions, as well as the role of (delocalized) high-frequency modes.

\subsection{Correlation functions and regularization}

The operators of the massless 1D scalar fields in the two cavities are
\begin{eqnarray}
\label{fieldop}
\phi (x_1) &=& \sqrt{\frac {\hbar c^2}{\Lo}} \sum_k \frac {\sin (k_k x_1)}{\sqrt{\wk}} \left( a_k + a_k^\dagger \right) ,
\\
\phi (x_2) &=& -\sqrt{\frac {\hbar c^2}{\Lo}} \sum_k \frac {\sin (k_k x_2)}{\sqrt{\wk}} \left( c_k + c_k^\dagger \right) ,
\label{fieldop1}
\end{eqnarray}
where $x_1$ is a point in the region $(0,\Lo )$ (left cavity) and $x_2$ in the region $(\Lo ,2\Lo )$ (right cavity). The possible values of $k_k$ are those allowed by the boundary conditions, that is
$k_k=n_k\pi /\Lo$, with $n_k=1,2,...$. In Eq. (\ref{fieldop1}), we have taken into account that $\sin [ k_k(2\Lo - x_2) ] = -\sin (k_kx_2)$.
All quantities in Eqs. (\ref{fieldop}) and (\ref{fieldop1}) are referred to the equilibrium position of the movable wall \cite{Law95}.

We aim to evaluate relevant correlations between field observables in the two cavities over the interacting ground state (\ref{corrected state}).
Using Eqs. (\ref{corr-1}), (\ref{corr-2}) and (\ref{corrected state}) and the expressions above of the field operators, it is immediate to see that
\begin{equation}
\label{fieldcorr}
\langle \tilde{g} \lvert \phi (x_1) \phi (x_2) \rvert \tilde{g} \rangle - \langle \tilde{g} \lvert \phi (x_1) \rvert \tilde{g} \rangle \langle \tilde{g} \lvert \phi (x_2) \rvert \tilde{g} \rangle = 0 .
\end{equation}
It is easy to see that the vanishing of the correlation function (\ref{fieldcorr}) is valid at any order of perturbation theory, as a consequence of the fact that the effective Hamiltonians (\ref{effint}) are quadratic in the field annihilation and creation operators. This means that the field operators in the two cavities are not correlated.

We can now evaluate the correlation of the squared fields, that is a very important local field observable, being also related to the field energy density and to the interaction of the field with test bodies (as we will discuss later on in this section)
\begin{equation}
\label{corrsqf}
C(x_1,x_2) = \langle \tilde{g} \lvert \phi^2(x_1) \phi^2(x_2) \rvert \tilde{g} \rangle - \langle \tilde{g} \lvert \phi^2(x_1) \rvert \tilde{g} \rangle \langle \tilde{g} \lvert \phi^2(x_2) \rvert \tilde{g} \rangle .
\end{equation}

After a lengthy algebraic calculation, we finally find
\begin{eqnarray}
\label{corrsqf2}
&\ & C(x_1,x_2) = -\frac {\hbar^3c^4}{\Lo^4m\wo} \sum_{pqrs} (-1)^{p+q+r+s}
\nonumber \\
&\times& \Big\{ \frac {\sin (k_p x_1)\sin (k_q x_1) \sin (k_r x_2) \sin (k_s x_2)}{(\wo +\omega_p +\wq)(\wo +\omega_r +\ws)}
\nonumber \\
&\ & \ + \Big[ \frac {\sin (k_p x_1)\sin (k_q x_1) \sin (k_r x_2) \sin (k_s x_2)}{(\wo +\omega_p +\wq)(\omega_p +\wq +\omega_r +\ws)}
\nonumber \\
&\ & \ \ + (x_1 \leftrightarrow x_2) \Big] \Big\} ,
\end{eqnarray}
where the first term comes from first-order processes with a pair of virtual photons emitted in one of the two cavities, while the second term (inside square brackets) comes from second-order processes with two pairs of virtual photons, one pair in each cavity.

In order to explicitly evaluate the correlation (\ref{corrsqf2}), we now consider the continuum limit by taking $\Lo \rightarrow \infty$, thus recovering the case of a single trembling mirror in the vacuum space separating two infinite half-spaces. In this limit, the sums over $k$ in (\ref{corrsqf2}) become integrals over the continuous variables $k_i=\omega_i/c, \; (i=p,q,r,s)$, by using $\sum_k \rightarrow (\Lo /2\pi)\int dk$; we also rescale these variables with respect to
$k_0 = \wo /c$: $k_p=\bar{p}k_0$, $k_q=\bar{q}k_0$, and $k_r=\bar{r}k_0$, $k_s=\bar{s}k_0$, with $\bar{p},\bar{q},\bar{r}, and\bar{s}$ dimensionless continuous variables running from $0$ to $\infty$.

In the continuum limit, the sums in Eq. (\ref{corrsqf2}) relative to the first cavity ($0 \leq x_1 \leq \Lo$) yield quantities of the following form
\begin{eqnarray}
\label{conlim1}
&\ &\sum_p (-1)^p \sin (k_p x_1 ) f(k_p) \nonumber \\
&\ & = \sum_p \cos (k_p \Lo) \sin (k_p x_1 ) f(k_p) ,
\end{eqnarray}
where $f(k_p)$ is a regular function and we have used $k_p = p\pi /\Lo$ ($p$ integer). If we then take $\sum_p \rightarrow (\Lo /2\pi )\int_0^\infty dk_p$, the quantity in Eq. (\ref{conlim1}) becomes
\begin{equation}
\label{conlim2}
\frac 12 \frac {\Lo}{2\pi}\int_0^\infty dk_p f(k_p) \left[ \sin (k_p (x_1+\Lo )) - \sin (k_p (\Lo - x_1)) \right] .
\end{equation}
Taking into account that the first term in the integral in Eq. (\ref{conlim2}) is rapidly oscillating at a finite distance from the movable wall, while the second term is slowly varying for $x_1 \sim \Lo$, we can neglect the first term and keep only the second one. Similar considerations apply also for the second cavity ($\Lo \leq x_2 \leq 2\Lo$).
We also introduce the two scaled (dimensionless) distances from the movable wall
\begin{equation}
\label{scaleddist}
d_1 = k_0 (\Lo -x_1) , \, \, d_2 = k_0 (x_2 -\Lo ) .
\end{equation}

After some algebra, the correlation function (\ref{corrsqf2}) thus becomes
\begin{eqnarray}
\label{corrsqf3}
&\ & C(d_1,d_2) = -\frac {\hbar^3c k_0}{2^4(2\pi )^4m}
\nonumber \\
&\times& \Big\{ \int \!\! \int \! d\bar{p} d\bar{q} \frac {\sin (\bar{p}d_1) \sin (\bar{q}d_1)}{1+\bar{p}+\bar{q}} \int \!\! \int \! d\bar{r} d\bar{s} \frac {\sin (\bar{r}d_2) \sin (\bar{s}d_2)}{1+\bar{r}+\bar{s}}
\nonumber \\
&\ & + \Big[ \int \!\!\int \!\!\int \!\!\int \!\!d\bar{p} d\bar{q} d\bar{r} d\bar{s} \frac {\sin (\bar{p}d_1) \sin (\bar{q}d_1)\sin (\bar{r}d_2) \sin (\bar{s}d_2)}{(1+\bar{p}+\bar{q})(\bar{p}+\bar{q}+\bar{r}+\bar{s})}
\nonumber \\
&\ & \ \ \ + (d_1 \leftrightarrow d_2) \Big] \Big\} .
\end{eqnarray}

All integrals in Eq. (\ref{corrsqf3}) extend from $0$ to $\infty$,
and an appropriate regularization scheme is necessary to deal with the high-frequency field modes. Also, even if our model considers the idealized case of a perfectly reflecting boundary, a regularization of the frequency integrals is necessary to take into account the properties of a real metal boundary, that becomes transparent for frequencies larger than the plasma frequency of the material. From a physical point of view, we however expect that the contribution of high-frequency modes becomes negligible at large distances from the boundary. We will address in more detail this important point, and its physical consequences in the case of a real metal, in the subsequent part of this paper.

We now introduce in (\ref{corrsqf2}) a regularization function for each $k_v$ integration ($v=p,q,r,s$) in the form of an exponential function $e^{-k_v/k_M}$, with $k_M$ a regularization scale factor. For a real metal, it can be set of the order of $\omega_p/c$, with $\omega_p$ the metal plasma frequency \cite{Novotny-Hecht12}. In the first term in the right-hand side of (\ref{corrsqf3}), the four integrals are factorized in two double integrals. Such integrals are in terms of the scaled variables previously introduced, and the exponential regularization function assumes the form $e^{-\mu \bar{v}} \, (\bar{v}=\bar{p},\bar{q},\bar{r},\bar{s})$ for each integration variable, with $\mu = (k_M/k_0)^{-1}$; the double integrals take then the form
\begin{equation}
\label{integral1}
I_\mu (d) = \int_0^\infty \! \! \int_0^\infty \! dr ds \frac {\sin (rd) \sin (sd)}{1+r+s} e^{-\mu r} e^{-\mu s} .
\end{equation}
The limit $\mu \rightarrow 0 \, \, (\omega_p \rightarrow \infty )$ restores the case of a perfectly reflecting boundary. This integral can be evaluated analytically, and, after some algebra and extensive use on integral tables \cite{Gradshteyn-Ryzhik15}, we obtain
\begin{eqnarray}
\label{integral4}
&\ & I_\mu (d) = \frac 14 \left[ e^{\mu - id} \text{$E_1$}(\mu -id) \left( \frac 1{id} +1 \right) \right.
\nonumber \\
&\ & \left. +  e^{\mu + id} \text{$E_1$}(\mu +id) \left( \frac {-1}{id}+1 \right) \right]
-\frac 12 \frac {\mu}{\mu^2+d^2} ,
\end{eqnarray}
where $E_1(z)$ is related to the analytic continuation of the exponential integral function to the complex plane, with $E_1(x)=-\text{Ei}(-x)$ for $x$ real.
 \cite{NISTHandbook10}. In the limit $\mu \rightarrow 0$ ($k_M \rightarrow \infty$, that is a perfect mirror), Eq. (\ref{integral4}) reduces to
\begin{equation}
\label{integral1a}
 I_{\mu =0}(d)= \frac 12 \left( \frac {f(d)}{d} + g(d) \right) ,
\end{equation}
where $f(x) = \text{ci}(x)\sin (x) + \text{si}(x)\cos (x)$ and $g(x) = -\text{ci}(x)\cos (x) - \text{si}(x)\sin (x)$ are the auxiliary functions of the sine-integral and cosine-integral functions, respectively \cite{NISTHandbook10}.

We are mainly interested in the case $d \gg 1$, that is, distances from the movable wall much larger than $c/\wo$. In this limit case, we can approximate Eq. (\ref{integral4}) using the asymptotic expansion of the exponential integral function \cite{Gradshteyn-Ryzhik15}, taking also into account that $\mu = \wo /\omega_M \ll 1$ for typical values of the mirror's oscillation frequency ($\wo \sim 10^5 \, \text{s}^{-1}$) and of the plasma frequency($\omega_M \sim \omega_p \simeq 1.4 \cdot 10^{16} \, \text{s}^{-1}$ for gold). For $d \gg 1$ and $\mu \ll 1$, using the asymptotic relation
$e^zE_1(z) \simeq \frac 1z$ \cite{NISTHandbook10}, we get
\begin{equation}
\label{integral4a}
I_\mu (d) \simeq \frac 12 \frac 1{\mu^2+d^2} \simeq \frac 1{2d^2}.
\end{equation}
Equation (\ref{integral4a}) also shows that the frequency integration, in the limits considered, is scarcely dependent from the ultraviolet cutoff $\mu$, and that the contribution of high-frequency modes to the integral is negligible, as indeed expected from physical grounds.

The integrals in the second term of Eq. (\ref{corrsqf3}) cannot be factorized, and after some algebraic manipulation the four frequency integrals, after introducing en exponential cutoff function, can be cast in the following form
\begin{eqnarray}
\label{integral2}
I_\mu (d_1,d_2) &=& \int_0^\infty \!\!\int_0^\infty \!\!\int_0^\infty \!\!\int_0^\infty \!\!dp dq dr ds e^{-\mu (p+q+r+s)}
\nonumber \\
&\ & \times \frac {\sin (pd_1) \sin (qd_1)\sin (rd_2) \sin (sd_2)}{(1+p+q)(p+q+r+s)} .
\end{eqnarray}
After some algebra, we finally obtain
\begin{widetext}
\begin{eqnarray}
\label{integral2a}
I_\mu (d_1,d_2) &=& \frac 18 \int_0^\infty \!\! dv \frac {e^{-\mu v}}{1+v} \left(  \frac {\sin (vd_1)}{d_1} -v \cos (vd_1)\right)
\left[ e^{v(\mu +id_2)} E_1(v(\mu +id_2)) \left( \frac {i}{d_2}+v\right) \right.
\nonumber \\
&\ & + \left. e^{v(\mu -id_2)} E_1(v(\mu -id_2)) \left( \frac {-i}{d_2}+v\right) \right]
-\frac 14 \int_0^\infty \!\! dv \frac {ve^{-\mu v}}{1+v} \left(  \frac {\sin (vd_1)}{d_1} -v \cos (vd_1)\right) .
\end{eqnarray}
\end{widetext}

In the limit $\mu \rightarrow 0$,
using $g(z) \pm if(z) = E_1(\mp iz) e^{\mp iz}$ \cite{NISTHandbook10},
Eq. (\ref{integral2a}) reduces to
\begin{eqnarray}
\label{integral2b}
I_{\mu = 0} (d_1,d_2) &=& \frac 14 \int_0^\infty \! \! \!\! dv \frac {v^2}{1+v} \left( \frac{\sin (vd_1)}{vd_1} - \cos (vd_1) \right)
\nonumber \\
&\ & \times \left( \frac {f(vd_2)}{vd_2} + g(vd_2) \right) .
\end{eqnarray}

The integral (\ref{integral2a}) must be evaluated numerically.
As before, we consider the case of large distances (compared to $k_0^{-1}$) from the movable wall, $d_1, \, d_2 \gg 1$, and cutoff frequency much larger than $ck_0$, that is $\mu \ll 1$. In this case, using the same asymptotic expansion of $E_1(z)$ used for Eq. (\ref{integral4}), we easily obtain
\begin{equation}
\label{integral5}
I_\mu (d_1,d_2) \simeq \frac 18 \frac 1{\mu^2+ d_1^2}  \frac 1{\mu^2+ d_2^2} \simeq \frac 1{8d_1^2d_2^2} .
\end{equation}
Similarly to the previous integral, this shows that, in the limits considered, the integral is quite independent from the cutoff frequency, and the contributions of high-frequency modes are negligible, consistently with our nonrelativistic treatment. The correlation function between the squared fields is thus
\begin{eqnarray}
\label{corrsf6}
C(d_1,d_2) &=& -\frac {\hbar^3c k_0}{2^4(2\pi )^4m} \left( I_\mu (d_1)  I_\mu (d_2) + I_\mu (d_1,d_2) \right)
\nonumber \\
&\ & \simeq  -\frac {\hbar^3c k_0}{2^5(2\pi )^4m} \frac 1{d_1^2d_2^2} .
\end{eqnarray}

In terms of nonscaled distances from the mobile wall $\tilde{x}_{1,2} = d_{1,2}k_0^{-1} = d_{1,2} c/\wo$, it becomes
\begin{equation}
\label{corrsf6a}
C(\tilde{x}_1,\tilde{x}_2) = -\frac {\hbar^3c^4}{2^5(2\pi )^4} \frac 1{m \wo^3}\frac 1{\tilde{x}_1^2 \tilde{x}_2^2}.
\end{equation}

Eq. (\ref{corrsf6a}) shows that the squared fields at the two sides of the reflecting mirror are anticorrelated, even if they are independent fields and there is not any direct interaction between them. This correlation is entirely ascribable to their mutual interaction with the movable wall, that allows a sort of influence between the two half-spaces, even in the vacuum state. The negative sign of $C(\tilde{x}_1,\tilde{x}_2))$ in Eq. (\ref{corrsf6a}) shows that the squared field are indeed anticorrelated; this sign can be understood from a physical point of view, by considering that the two field-mirror coupling constants $C_{kj}^1$ and $C_{kj}^2$ have an opposite sign, and thus when the position of the mirror fluctuates in one direction, the modes of one cavity are blueshifted while those of the other cavity are redshifted.

If we take the two points considered at the same distance from the mirror's equilibrium position on the opposite side, $\tilde{x}_1=\tilde{x}_2=x$, Eq. (\ref{corrsf6a}) yields that at large distance from the movable wall the (anti)correlation between the squared field scales with the distance as $x^{-4}$. It also depends as $\frac 1{m\wo^3}$ from the mass and oscillation frequency of the movable wall. This means that the smaller the mirror's mass and oscillation frequency are, the larger the effect we find is. In modern optomechanics experiments, a typical oscillation frequency is of the order of $10^5$ Hz, but also higher frequencies can be obtained, and a mass as low as $10^{-21}$ Kg can be achieved \cite{Aspelmeyer-Kippenberg14}. However, we wish to stress that our results above show that a correlation between the squared fields at the opposite sides of the movable mirror exists at any distance from it.

Using Eqs. (\ref{integral1}), (\ref{integral4}), (\ref{integral2}), and (\ref{integral2a}) in Eq. (\ref{corrsqf3}), we obtain the expression for the correlation function for points at any distance from the movable mirror, with the only condition, as discussed later on, $\tilde{x}_{1,2} \gg c/ \wpl$,
\begin{eqnarray}
\label{corrsqf4}
C_\mu (d_1,d_2) &=& -\frac {\hbar^3c k_0}{2^4(2\pi )^4m} \left[ I_\mu (d_1) I_\mu (d_2) \right.
\nonumber \\
&\ & \left. + I_\mu (d_1,d_2) + I_\mu (d_2,d_1)\right] ,
\end{eqnarray}
where in the left-hand-side term we have explicitly indicated the dependence on the regularization parameter $\mu$. The integral in (\ref{integral2a}) can be obtained numerically.

\subsection{Observability through dispersion interactions}

Observability of the spatial correlation of the squared field could be obtained exploiting the dispersion interaction of the fluctuating field with a polarizable object $A$, that in the far-distance (retarded) regime is usually proportional to the square of the field at the polarizable-body position $r_A$, in the form
\begin{equation}
\label{interpol}
\Delta E = -\frac 12 \alpha_A \langle \phi^2(r_A) \rangle ,
\end{equation}
where the average value is taken on the dressed ground state of the system and $\alpha_A$ is the static polarizability of the polarizable body $A$ \cite{Passante18}.
The relation above can be applied in our case at a sufficiently large distance from the wall, where only low-frequency field modes are relevant, as we have shown; this is totally consistent with the well-known results for the far-zone (retarded regime) Casimir-Polder dispersion interaction between atoms or between an atom and a conducting surface, where the dynamical polarizability can be replaced with its static value, because only low-frequency modes contribute in the far zone. In fact, in this case for $x \gg c/\omega_A$ (with $\omega_A$ a main transition frequency of the atom or polarizable body), only field modes with $\omega \ll \omega_A$ give a significant contribution to the dispersion interaction energy, while the contribution of higher-frequency modes is negligible \cite{Power-Thirunamachandran92,Passante18}

The effect we find should therefore yield a correlation between the dispersion interaction between the mirror and two polarizable bodies $A$ and $B$ (ground-state atoms or molecules, for example, in the electromagnetic case), placed at the opposite sides of the movable reflecting wall, as well as an atom-atom dispersion interaction mediated by the trembling mirror. A joint measurement of the dispersion force on $A$ and $B$, and their correlation as a function of the atom-mirror distance, should allow one to observe the spatial correlation between the squared field in Eqs. (\ref{corrsf6}) and (\ref{corrsf6a}). This is also motivated by the fact that the movable mirror acts as a third body interacting with the field, and thus it is capable of modifying the radiation-mediated interacting energy between the two polarizable bodies $A$ and $B$ \cite{Passante-Persico99,Milton-Abalo13}.  Another possible experimental observation could result from three-body effects on the atom-atom dispersion interaction between two atoms (or, in general, polarizable bodies) $A$ and $B$ at one side of the wall due to a third atom $C$ placed on the other side of the wall. In fact, the three-body component of the dispersion force on one atom, say atom $A$, is related to a sort of interference between the virtual field energy densities due to the two other atoms $B$ and $C$ \cite{Passante-Persico99} or to the virtual field correlation at the location of $A$ and $B$, dressed by $C$ \cite{Cirone-Passante97}. For a fixed wall, atom $C$ cannot influence the interaction between $A$ and $B$, while in the present case of a movable wall it can: thus, a measurement of the $A$-$B$ interaction should keep trace of the presence of $C$, even if it is placed on the opposite side of the perfectly reflecting movable wall.

\subsection{Contribution from high-frequency modes}

As we mentioned in the introduction, in the case of a static perfect mirror the two half-spaces are totally separated and uncorrelated, while a (anti)correlation between the squared fields in the two half-spaces arises if the mirror is allowed to move (even in the ground state).

Even if our main point is the conceptual aspect that the two half-spaces can influence each other even for an ideal mirror, if it is allowed to move, we wish to make here some general considerations about the role of high-frequency modes for a real static mirror, that becomes partially transparent at high frequencies (larger than its plasma frequency). In the case of a real static mirror, there is thus some influence between the two half-spaces, both for the field and the squared field, that involves only high-frequency field modes, specifically those above the metal plasma frequency $\wpl$. This effect can add up to the effects we found for the movable mirror, the latter essentially due to low-frequency modes. We wish here to discuss the consequence of this consideration for our new results for the movable perfect mirror.

For very high-frequency modes, $\omega \gg \wpl$, the mirror separating the two cavities becomes transparent and these modes are delocalized in the two cavities. The scalar field operator is then approximately
\begin{equation}
\label{fieldhf}
\phi (x) = \sum_{k > \wpl /c} \left( \frac {\hbar c^2}{2L\wk} \right)^{1/2} \left( a_k e^{ikx} + a_k^\dagger e^{-ikx} \right) ,
\end{equation}
where $L=2L_0$ is the distance between the two fixed cavity walls, and the sum must be taken only over the high-frequency modes, that is with frequencies higher than a typical metal plasma frequency $\omega_{pl}=ck_{pl}$; they must also satisfy the boundary conditions at the two fixed walls. A simple calculation yields the following correlation function on the vacuum state $\lvert 0 \rangle$
\begin{widetext}
\begin{eqnarray}
\label{corrff}
C_{f-static} (\Delta x = x-x') &=& \langle 0 \lvert \phi (x) \phi (x') \rvert 0 \rangle - \langle 0 \lvert \phi (x) \rvert 0 \rangle \langle 0 \lvert \phi (x') \rvert 0 \rangle = \sum_{k > \wpl /c} \frac {\hbar c^2}{2L\wk} e^{ik(x-x')}
\nonumber \\
&\ & = \frac {\hbar c}{2\pi} \int_{k_{pl}}^\infty \!\! dk \frac {\cos (k(x-x'))}{k} = - \frac {\hbar c}{2\pi} \text{Ci} (k_{pl}(x-x')) ,
\end{eqnarray}
\end{widetext}
where in the second line the continuum limit $L \rightarrow \infty$ has been taken, and $\text{Ci}(z)$ is the cosine integral function \cite{NISTHandbook10}. We will discuss in more detail the meaning of Eq. (\eqref{corrff}) in the following of this section, after evaluation of the correlation of the squared fields.

We now evaluate the free-space correlation of the squared field, due to the delocalized high-frequency modes with $\omega > \wpl$, and compare their role with the results previously obtained when the movable plate is present
\begin{widetext}
\begin{eqnarray}
\label{corrsff}
C_{sf-static} (\Delta x = x-x') &=& \langle 0 \lvert \phi^2 (x) \phi^2 (x') \rvert 0 \rangle - \langle 0 \lvert \phi^2 (x) \rvert 0 \rangle \langle 0 \lvert \phi^2 (x') \rvert 0 \rangle
= 2 \left( \frac {\hbar c^2}{2L} \right)^2 \left[ \sum_{k > \wpl /c} \frac {1}{\wk} e^{ik(x-x')} \right]^2 .
\nonumber \\
&\ &
\end{eqnarray}
\end{widetext}
In the continuum limit $L \rightarrow \infty$ (free space), we get
\begin{equation}
\label{corrsffc}
C_{sf-static} (\Delta x = x-x') = 2 \left( \frac {\hbar c^2}{2L} \right)^2 \text{Ci}^2 (k_{pl}(x-x')) .
\end{equation}
The integrals (\ref{corrff}) and (\ref{corrsffc}) do not contain the cutoff function because they are relative to an unbounded free space, and, moreover, it should be noted that they converge, except for $x=x'$ (regime not relevant in our case, since we are interested only in points such that $k_{pl}(x-x') \gg 1$).

We now analyze the results for the unbounded-space correlations (\ref{corrff}) and (\ref{corrsffc}) (contribution of extremely high-frequency modes where any real metal becomes transparent), in comparison with our main results (\ref{fieldcorr}), (\ref{corrsqf3}) and (\ref{corrsf6a}) in the presence of the fluctuating movable wall. Using the asymptotic expansion of the sine integral function, $\text{Ci}(y) \sim \cos(y)/y$ for $y \gg 1$
\cite{NISTHandbook10}, Eqs. (\ref{corrff}) and (\ref{corrsffc}) for $\lvert x-x' \rvert \gg k_{pl}^{-1}$ reduce to
\begin{eqnarray}
\label{correfasymp}
C_{f-static} (x-x') &\simeq& - \frac {\hbar c}{2\pi} \frac {\cos [k_{pl}(x-x')]}{k_{pl}\lvert x-x' \rvert}
\\
C_{sf-static} (x-x') &\simeq 2& \left( \frac {\hbar c}{2\pi} \right)^2 \frac {\cos^2 [k_{pl}(x-x')]}{[k_{pl}( x-x')]^2} .
\label{correfsasymp}
\end{eqnarray}
Equation (\ref{correfasymp}) shows that, in the case of an imperfect static mirror, there is a nonvanishing vacuum correlation between the fields at the two sides on the mirror, as indeed expected by simple physical considerations, that however goes to zero for high cutoff frequency and/or large distance between the two points considered with fast space oscillations yielding a vanishing average, while the contribution (\ref{fieldcorr}) of the fluctuating motion of the movable mirror is zero, within our approximations. However, what is more relevant is the comparison between the correlation of the squared field at large distances, i.e. Eq. (\ref{corrsf6a}) when the movable mirror is present, and Eq. (\ref{correfsasymp}) in the case of a fixed imperfect mirror. This latter comparison immediately shows that the two expressions scale with a different power of the unscaled distance (for large distances), that is ${\tilde{x}}^{-4}$ and ${\tilde{x}}^{-2}$ respectively. Also, they have opposite sign. We can evaluate their ratio $\lvert C(\tilde{x},\tilde{x}) \rvert /C_{sf-static}(2\tilde{x})$, for simplicity at two points at the same distance $\tilde{x}$ for the movable mirror, with typical numerical values of the parameters involved,
$m$, $\wo$, and $\wpl$.
From Eqs. (\ref{corrsqf4}) and (\ref{correfsasymp}), we find the following expression of the ratio above
\begin{equation}
\label{ratiosf}
\frac {\lvert C(\tilde{x},\tilde{x}) \rvert}{C_{sf-static}(2\tilde{x})} \simeq \frac {\hbar \wpl^2}{2^4 \pi^2 m c^2 \wo} d^2 \left[ (I_\mu (d))^2 + 2I_\mu (d,d) \right] ,
 \end{equation}
with $d=\tilde{x}\wo /c$, and the values of the functions $I_\mu (d) and I_\mu (d,d)$ can be in general obtained numerically. Its value depends on the mass, oscillation and plasma frequency of the movable wall, as well as from the distance.
The smaller $m$ and $\wo$, the larger the ratio (\ref{ratiosf}) is. As an example, if we set $\wo = 10^4 \, \text{s}^{-1}$, $\wpl = 1.5 \cdot 10^{16} \, \text{s}^{-1}$ (that is, the plasma frequency of silver), we find that for $m=10^{-23}$ Kg there exists a large range of the scaled distance $d$, with $\tilde{x} \gg c/\wpl$ and $\tilde{x} \ll c/\wo$, for which the contribution to the correlation due to the movable wall is comparable in size with the static one due to the high-frequency modes for which the movable wall becomes transparent. Also, the numerical evaluation of the integrals in $I_\mu (d)$ and $I_\mu (d,d)$ shows that these quantities do not depend significantly from the cutoff frequency $\wpl$ if $\wpl \gg \wo$ and $\tilde{x} \gg c/\wpl$, condition very well verified in any realistic setup; thus, the high-frequency field modes have a negligible role to the correlation $C_\mu (d,d)$ induced by the movement of the wall, coherently with our approach.
In other words, the effect of the trembling wall on the squared field spatial correlations can exceed that of the modes above the conductor's plasma frequency for which the wall becomes transparent, with appropriate choices of the relevant parameters of the movable wall; moreover, the sign of the correlation function due to the trembling can be negative (anticorrelation), while the "static" contribution of high-frequency modes is always positive. However, we wish to stress the conceptual relevance of our results showing that, even for a perfect conductor, the wall's quantum position fluctuations allow the existence of correlations between field observables at the opposite sides of the boundary. Generalization to the more realistic case of the quantum electromagnetic field in a three-dimensional space will be considered in a future work.

Finally, we wish to mention that our results have some resemblance to the case of Hawking radiation from a fluctuating event horizon due to quantum gravity effects, where a fuzzy event horizon of a black hole is present and radiation is emitted by the fluctuating boundary \cite{Arias-Krein12,Takahashi-Soda10}; also, our results might indicate the possibility that a fluctuating event horizon could allow a correlation between physical observables defined in the internal and external regions of the horizon, or across the cosmological event horizon for an expanding Universe. We hope to investigate this intriguing aspect in the near future.

\section{\label{sec:4}Conclusion}

We have considered a system of two cavities separated by a reflecting mirror that can fluctuate around its equilibrium position, with two massless one-dimensional scalar fields defined in the two half-spaces separated by the movable wall.
Assuming small displacements of the movable wall from its equilibrium position, an effective field-mirror interaction exists, as well an effective interaction between the field modes mediated by the movable mirror; this generates changes in global and local field quantities in the interacting ground state, with respect to the fixed-wall case. A previous work has shown that some quantities such as time-dependent energy shifts, as well as field energy densities, do not display influence between the two cavities up to the second order in the wall-field effective coupling \cite{Armata-Kim17}. In this paper, we have focused our attention to correlation functions between field observables in the two cavities. We have calculated the interacting ground state of this system at the second order of the field-mirror effective interaction, that contains both field and mirror virtual excitations. We have shown that
although the ground-state correlation function between the field operators at opposite sides of the movable wall vanishes,
a nonvanishing anticorrelation between the squared fields evaluated at the two sides of the mirror exists.
We have evaluated its dependence on the distance from the mirror's equilibrium position and its scaling from the mass and oscillation frequency of the movable mirror. This clearly shows a sort of communication between the two half-spaces, mediated by the fluctuating boundary, even in the case of a perfectly reflecting mirror. Possible observability of this new effect has been also discussed, showing that in principle it can measured exploiting the retarded Casimir-Polder interactions between polarizable bodies placed at the opposite sides of the movable wall.
We have also shown that our results, at large distances from movable wall, are quite insensitive to the ultraviolet cutoff frequency, and discussed also the case of a real conductor characterized by its plasma frequency.
We stress that these new results could be relevant also in different physical systems, such as quantum fields at the two sides of a fluctuating event horizon of a black hole in quantum gravity theories. Generalization to the three-dimensional case and to the quantum electromagnetic field case will be the subject of future publications.

\section*{Acknowledgements}
The authors gratefully acknowledge financial support from the Julian Schwinger Foundation and Ministero dell'Universit\`{a} e della Ricerca (MUR). R.P. and L.R. also acknowledge partial financial support from the FFR2021 grant from the University of Palermo, Italy.

\bibliography{biblio}

\end{document}